5 pages, 4 figures, no macros
\documentstyle[pre,aps]{revtex}
\begin{document}
\draft
\title{Time evolution of near membrane layers}

\author{Kazimierz Dworecki\footnote{Electronic address:
dworecki@pu.kielce.pl}, 
Tadeusz Koszto\l owicz\footnote{Electronic address:
tkoszt@pu.kielce.pl},
S\l awomir W\c asik}
 
\address{Institute of Physics, Pedagogical University,\\
ul. Konopnickiej 15, PL - 25-406 Kielce, Poland}

\author{Stanis\l aw Mr\' owczy\' nski\footnote{Electronic address:
mrow@fuw.edu.pl}}

\address{Institute of Physics, Pedagogical University,\\
ul. Konopnickiej 15, PL - 25-406 Kielce, Poland\\
and So\l tan Institute for Nuclear Studies,\\
ul. Ho\.za 69, PL - 00-681 Warsaw, Poland}

\date{1-st February 1999}
 
\maketitle

\begin{abstract}

The near membrane layer is defined as a region where the concentration 
of the substance transported across the membrane drops $k$ times. The time 
evolution of such a layer is studied experimentally by means of the laser 
interferometric method. It is shown that within the experimental errors 
the thickness of the near membrane layer grows in time as $a \sqrt t$ 
with the coefficient $a$ being independent of the initial concentration and 
the membrane permeability. Time evolution of the near membrane layers is 
also analyzed within the theoretical approach earlier developed by one 
of us. The regularities found experimentally fully agree with the 
theoretical expectations.

\end{abstract}

\vspace{0.5cm}
PACS number(s): 66.10.Cb, 82.65.Fr 

\vspace{1cm}

\section{Introduction}

The transport in membrane systems is of great interest in several fields 
of technology \cite{Rau89}, where the membranes are used as filters, 
and biophysics \cite{Tho76}, where the membrane transport plays a crucial 
role in the cell physiology. The diffusion in a membrane system 
is also interesting by itself as a nontrivial stochastic problem. While the 
time dependent concentration profiles of the substance transported across the 
membrane give a detailed description of the macroscopic substance motion, 
we are often interested only in the regions with a sufficiently large 
concentration. Such a situation occurs when the phenomenon under consideration 
strongly depends on the concentration. For example, the hydrodynamic stability 
in the membrane systems studied in \cite{Sle85} is controlled by the Rayleigh 
number, which in turn depends on the transported substance concentration 
\cite{Cro93}. Therefore, it is sometimes convenient to introduce the so-called 
{\it near membrane layer} (NML) where the substance concentration drops $k$ 
times \cite{Ler76}. In this paper we study NML experimentally and 
theoretically. 

Our experimental investigation is carried out by means the laser 
interferometric method. The laser light is spilt into two beams. The first 
one goes through the membrane system parallelly to the membrane surface while 
the second, reference one goes directly to the light detecting system. The 
interferograms, which appear due to the interference of the two beams, are 
controlled by the refraction coefficient of the solute which in turn depends 
on the substance concentration. The analysis of the interferograms allows one 
to reconstruct the time dependent concentration profiles of the substance 
transported across the membrane. Further one can find how NML evolve in time. 
We show that the time evolution of the NML thickness manifests surprisingly 
simple regularities.

The time dependence can be studied using the Smoluchowski (diffusion) 
equation. However, one has to impose two boundary conditions at the membrane 
surface. The first one is provided by the substance current conservation 
but there is {\it no} obvious choice of the second condition. When the 
membrane has a finite thickness, the diffusive transport within the membrane 
is often described by the Smoluchowski equation as well \cite{Hoo76}. The 
diffusion constant however differs from that one which is in the regions 
outside the membrane. Then, one assumes \cite{Hoo76} that the ratio of the 
concentrations at both sides of each of the membrane surfaces equals a 
constant which is a free parameter. We find this approach as not very 
satisfactory. Since a real membrane is not homogeneous and its internal 
structure is rather complicated, using the diffusion equation inside 
the membrane is rather questionable. The boundary condition, which fixes  
the concentration ratios, is introduced without physical justification. 
When the membrane thickness goes to zero, the membrane selectivity vanishes 
entirely, and consequently, the approach is hardly useful for very thin 
membranes. In the series of papers \cite{Kos96,Kos98a,Kos98b}, one of us 
has developed the approach which is also based on the Smoluchowski equation. 
The boundary condition is well motivated and the membrane can be treated 
as an infinitely thin wall characterized by the permeability coefficient. 
The time dependence of NML can be easily derived and, as we show here, 
it agrees very well with the experimental data.

The paper is organized as follows. In Sec. II we present the experimental
procedure with the results on the concentration profiles and the near membrane
layers. Sec. III is devoted to the theoretical considerations. The solution
of the Smoluchowski equation is found and the time dependence of the NML 
thickness is derived. The predictions of our theoretical model are compared
with the experimental data. We summarize our study in Sec. IV.

\section{Experiment}

The membrane system under study is the cuvette of two chambers 
separated by the horizontally located membrane. Initially we fill 
the upper (lower) chamber with the aqueous solution of the ethanol 
while in the lower (upper) one there is pure water. Then, the ethanol 
diffuses to the lower (upper) chamber.  Since the concentration 
gradients are in the vertical direction only, the diffusion is expected 
to be one dimensional (along the axis $x$). In other words, the ethanol 
concentrations are assumed to be uniform in the planes parallel to the 
membrane.  

As already mentioned we employ the laser interferometric method to measure
the time dependent concentration profiles in the membrane system. Let us note 
that the measurement does not disturb the system under study. The experimental 
setup is described in \cite{Dwo95}, here we only mention that it consists 
of the measurement cuvette with the membrane, the Mach-Zehnder interferometer 
\cite{Ste83} including the He-Ne laser, TV-CCD camera, and the computerized 
data acquisition system. 

The interferograms are sensitive to the variation of the refraction 
coefficient within the membrane system and consequently to the concentration 
gradients. When the solute is uniform the interference fringes are straight 
and they bent when the concentration gradient appears. The example of the 
interference images are shown in Fig. 1. The substance concentration at $x$ 
is determined by the deviation $d$ of the fringes from their straight 
line run. Since the relation between the concentration $C$ and the refraction 
coefficient is assumed to be linear, we have
$$
C(t,x) = C_0 + \alpha \, {\lambda \, d(t,x) \over h \, f} \;,
$$
where $C_0$ is the initial substance concentration; $a$ is the 
proportionality constant between the concentration and the refraction 
index, $\alpha =3.19 \cdot 10^5 \; {\rm mol}/{\rm m^3}$ for the ethanol aqueous 
solution; $\lambda$ is wavelength of the laser light; $h$ denotes the 
distance between the fringes in the field where they are straight lines; 
$f$ is the thickness of the solution layer in the measurement cuvette. 
Recording the interferograms with a given time step one can reconstruct 
the time dependent concentration profiles. 

We measured the profiles for several values of the initial ethanol 
concentration 
and for two cellulose membranes of different permeabilities. The membrane 
thickness is, respectively, 0.17 mm and 0.01 mm. In Fig. 2 we present the 
concentration profiles taken at different moments of time.  Repeating several 
times the measurements at the same conditions we tasted the stability of our 
results. While the shape of the concentration profiles has appeared to be 
rather stable, 
the absolute normalization has varied by about 15 \%. The error bars shown 
in Fig. 2 
just correspond to this uncertainty. 

Since the specific weight of the ethanol is significantly lower than that 
of water, there can appear a convective motion in the system when the ethanol 
is initially in the lower chamber. To check whether the gravitational force 
significantly influences the process we have studied the ethanol transport from 
the upper chamber to the lower one and then the opposite configuration. We have 
found that for sufficiently low initial ethanol concentrations under study 
(125 -- 750 mol/m$^3$), the two configurations are within the experimental 
uncertainties equivalent to each other. 

Having the profiles one can define the near membrane layer (NML).
When the substance diffuses across the membrane into the pure solvent, 
the thickness $l$ of NML is defined as a length at which the concentration 
decreases $k$ times i.e.
\begin{equation}\label{thickness}
C(t,x=0) = k \; C(t,x=l) \;,
\end{equation}
with $x=0$ being the membrane position\footnote{If the substance diffuses 
to the region of the nonzero initial concentration $C_0$, the definition 
(\ref{thickness}) is generalized as 
$C(t,0) - C_0 = k \big[ C(t,l) - C_0 \big].$ This form can be applied 
to the layer on both sides of the membrane.}.

Taking $k=12.5$ we found the thickness of NML from the earlier obtained 
concentration profiles. In Figs. 3 and 4 we present the thickness as 
a function of time. As seen it manifests the remarkable properties. The 
NML thickness appear to be independent, within the experimental errors, 
of the initial concentration (Fig. 3) and of the membrane permeability 
(Fig. 4). The thickness grows in time as $\sqrt{t}$. As we have already
mentioned our concentration profile measurements suffer from the absolute
normalization uncertainty of about 15 \%. However, this uncertainty does
{\it not} influence the NML thickness because the absolute normalization
coefficient drops out entirely in the NML defintion (\ref{thickness}).
Therefore, the estimated errors of the data points shown in Figs. 3 
and 4 are contained within the data point symbols. In the next section 
we discuss the NML thickness from the theoretical point of view.

\section{Theory}

Let us briefly present the approach to the membrane transport developed 
in \cite{Kos96,Kos98a,Kos98b}. The concentration profile $C(t,x)$ is 
assumed to satisfy the Smoluchowski equation 
\begin{equation}\label{smoluch}
{\partial C \over \partial t} = D {\partial^2 C \over \partial x^2} \;,
\end{equation}
where $D$ is the diffusion constant. As already mentioned we have 
experimentally checked that the we deal with a pure diffusion. Therefore,
in contrast to our earlier study \cite{Kos96}, the term responsible for the 
convection is neglected in (\ref{smoluch}). The membrane is treated as an 
infinitely thin, partially permeable wall. Then, one needs two boundary 
conditions at the wall to solve eq. (\ref{smoluch}). The first one is provided 
by the substance current conservation. The second one is given by the relation 
\begin{equation}\label{boundary1}
J(t,x=0) = (1 - \delta)\, J^0(t,x=0) \;,
\end{equation}
where $J(t,x)= - D\,\partial C/\partial x$ is the substance current in 
the membrane system while $J^0(t,x)=- D\,\partial C^0/\partial x$ denotes 
the current in the identical system but with removed membrane; $\delta$ is the 
membrane permeability coefficient, $0 \leq \delta \leq 1$. 

As already mentioned, the solute concentration is initially zero in one 
chamber of the membrane system and it is finite and uniform in the other 
one i.e.
\begin{displaymath}
C(t=0,x) = \left\{ \begin{array}{ccl} 
C_0 \;\;\; & {\rm for} &\;\; x < 0 \;, \\ 
0 \;\;\;\; & {\rm for} &\;\; x > 0 \;. \\
\end{array} \right. 
\end{displaymath}
The solution of eq. (\ref{smoluch}) for the given boundary and initial 
conditions reads
\begin{equation}\label{solution-}
C(t,x) = C_0 \bigg[1 - {1 - \delta \over 2}\, 
{\rm erfc}\Big(-{x \over 2 \sqrt{Dt}}\Big) \bigg] 
\;\;\;\;\;{\rm for}\;\; x < 0 \;,
\end{equation}
and 
\begin{equation}\label{solution+}
C(t,x) = C_0 {1 -\delta \over 2}\, {\rm erfc}\Big(
{x \over 2 \sqrt{Dt}}\Big) 
\;\;\;\;\;{\rm for}\;\; x > 0 \;,
\end{equation}
with ${\rm erfc}(x)$ being the complementary error function defined as
$$
{\rm erfc}(x)= {2 \over \sqrt{\pi}}\int_x^{\infty}dt \, e^{-t^2} \;.
$$

Substituting the solution (\ref{solution+}) into the layer thickness
definition (\ref{thickness}) we get
\begin{equation}\label{l-theor}
l(t) = a \, \sqrt{t} \;,
\end{equation}
where the coefficient $a$ depends solely on $D$ and $k$ as
$$
a=2 \sqrt{D}\,{\rm erfc}^{-1}\Big({1 \over k}\Big) \;,
$$
and ${\rm erfc}^{-1}(0.08)\cong 1.24$. One sees that the relation 
(\ref{l-theor}) fully agrees with the experimental results shown in Figs. 
3 and 4. Indeed, the dependence on the initial concentration $C_0$ and 
the membrane permeability coefficient $\delta$, which is present in the 
solution (\ref{solution+}), drops out entirely in eq. (\ref{l-theor}).

The near membrane thickness (\ref{l-theor}) is independent of $C_0$ 
due to the linearity of the diffusion equation(\ref{smoluch}). The 
cancellation of the membrane permeability in eq. (\ref{l-theor}) is 
much less trivial. A solution of the diffusion equation (\ref{smoluch}) 
usually depends on $x^2/Dt$, which is the only dimensionless 
combination of $D$, $x$ and $t$. Then, any length must be 
proportional to $\sqrt{Dt}$ as in our eq. (\ref{l-theor}).  However, 
the diffusion coefficient $D$ is {\it not} the only one dimensional 
parameter in the problem under consideration. For example, the 
membrane thickness $d$ can play such a role. Then, we can construct 
several dimensionless combinations of $x$ and $t$ e.g. $xd/Dt$. 
In our model the membrane is infinitely thin, so $d$ is irrelevant. 
However, the dimensional parameter can be introduced to the problem 
through the boundary condition. For example, the dimensional membrane 
permeability parameter $\chi$ appears in the condition of the form
$$
J(t,x=0) = \chi \Big[ C(t,x=0^+) - C(t,x=0^-)\Big] \;.
$$
Then, the solution of the diffusion equation, which satisfies this boundary 
condition, depends not only on $x^2/Dt$ but on other dimensionless 
combinations such as $\chi t/x$. Then, the thickness of the near membrane 
layer is $\chi$ dependent. Thus, one sees that our formula  (\ref{l-theor}) is 
independent of the membrane permeability because of the specific choice of 
the boundary condition (\ref{boundary1}). 

In our previous studies \cite{Dwo95,Kos96} we took the diffusion coefficient 
form the literature to fit the data. However, the coefficient is known to be 
sensitive to the temperature and the ethanol concentration in water. One sees
that our formula (\ref{l-theor}) solely depends on the diffusion coefficient.
Therefore, $D$ can be obtained directly from our data (shown in Figs. 3 and 4)
by means of a single parameter fit. In this way we get 
$D=0.76 \cdot 10^{-9} \;{\rm m^2/s}$. Since the value of $D$ is known, 
there is only one parameter $\delta$ to be fitted when the solution
(\ref{solution+}) is compared with the experimental concentration 
profiles. The resulting theoretical curves are shown in Fig. 2. The 
agreement with the experimental data is reasonably good but a comment
is in order here. 

Our theoretical approach assumes that the membrane is infinitely thin while 
the membranes which have been used in the measurements are, obviously, of 
the finite 
thickness. Further, there is a near membrane dead zone where the concentration 
measurement is unreliable or even impossible. The dead zone appears due to 
the imperfection of the cuvette edge and the small deformations of the membrane 
during the measurements. The latter effect, which is much more important for 
the thinner membrane, is not very well controlled in our set-up and it leads 
the 
substance stirring in the very vicinity of the membrane surface.  To take into 
account all these effects we have shifted the theoretical curves from Fig. 2 by 
0.3  mm to the left. This is, of course, not a quite satisfactory procedure and in 
the future studies we intend to resolve the problem in a systematic way. The main 
objective of this study, however,  are not the concentration profiles but the  near 
membrane layers. We have carefully checked that the time dependence of the 
near membrane layer is influenced insignificantly when the solution (\ref{solution+}) 
is shifted by 0.3 mm.

\section{Final remarks and summary}

The solution of the diffusion equation found with the boundary condition 
(\ref{boundary1}) is identical with the solution which satisfies the boundary
condition
\begin{equation}\label{boundary2}
C(t, x=0^-)=\kappa \, C(t,x=0^+)
\end{equation}
when  $\kappa$ is related to $\delta$ as 
$$
\kappa = {1 - \delta \over 1 + \delta} \;.
$$
Therefore, the formula (\ref{l-theor}) can be obtained within the boundary 
condition (\ref{boundary2}) which is well known in the diffusion theory.
However, as far as we know the time dependence (\ref{l-theor}) has not 
been earlier discussed. 

We also mention that the boundary condition (\ref{boundary1}) applies 
for a pure diffusion i.e. when the convection term is neglected in 
eq. (\ref{smoluch}). However, when the convection is present in the system, 
one easily generalizes eq. (\ref{boundary1}) \cite{Kos96,Kos98a,Kos98b} 
and introduce the parameters which separately determine, as in the 
Kedem-Katchalsky approach \cite{Kat65}, the membrane permeability 
with respect to the diffusion and to the convection. On the other hand, 
we see no way to generalize eq. (\ref{boundary2}) to distinguish
between the diffusive and convective membrane permeability. Therefore, 
we would like to stress that the equivalence of eqs. (\ref{boundary1}) 
and (\ref{boundary2}) holds only for a pure diffusion regime.

Let us summarize our study. Using the interferometric method the 
concentration profiles have been measured for the aqueous solution of the 
ethanol diffusing across the membrane into pure water. The time evolution 
of the near membrane layer has been then analyzed. The thickness of the layer, 
which is defined as a length where the concentration drops $k$ times, 
appears to grow in time as $a\sqrt{t}$ with the proportionality coefficient 
$a$ being remarkably independent of the initial ethanol concentration 
and of the membrane permeability. While the independence of $C_0$ is
the result of the problem linearity, the origin of the cancellation of the membrane 
permeability is much less trivial. Theoretical analysis performed within 
the approach proposed by one of us has shown that $a$ depends only on $k$ 
and the diffusion constant $D$ due to the specific choice of the boundary 
condition at the membrane. We have used this fact to obtain the numerical value 
of $D$ fitting the experimental time dependence of the near membrane layer. 
Then, the experimentally found concentration profiles have been fitted by the 
solution of the Smoluchowski equation. The membrane permeability coefficient 
$\delta$ is then a single fit parameter. 

\begin{acknowledgments}
We are very grateful to Konrad Bajer for fruitful discussions and
stimulating criticism. This work was partially supported by the 
Polish Committee of Scientific Research under Grant No. 2 P03B 129 16.
\end{acknowledgments}

\newpage
\vspace{2cm}
\begin{center}
{\bf Figure Captions}
\end{center}
\vspace{0.3cm}

\noindent
{\bf Fig. 1.} 
The interferograms which are analyzed to obtain the concentration profiles.
There is initially uniform ethanol solution of the concentration 125 
mol/m$^3$ in the upper part of the measurement cuvette. The interferograms
are taken at several values of time: (a) -- 240 s, (b) -- 600 s, and
(c) -- 1200 s.

\vspace{0.5cm}

\noindent
{\bf Fig. 2.} The concentration profiles for $C_0 = 125$ mol/m$^3$ taken 
at several values of time: $\circ$ -- 240 s, $\triangle$ -- 600 s, and
$\Box$ -- 1200 s.  The permeability of the membrane from (a) is significantly
smaller than that from (b). The solid lines represent the Smoluchowski 
equation solution with $D=0.76 \cdot 10^{-9} \;{\rm m^2/s}$ and 
$\delta = 0.51$ for (a) and $\delta = 0.11$ for (b). The theoretical 
curves are shifted to the left in (a) and (b) by 0.3 mm (see text).

\vspace{0.5cm}

\noindent
{\bf Fig. 3.} 
Time evolution of the near membrane layer for several values of the
initial ethanol concentrations: $\circ$ -- 125 mol/m$^3$, $\triangle$ -- 
250 mol/m$^3$, $\Box$ -- 500 mol/m$^3$, and $\star$ -- 750 mol/m$^3$.
The solid line represents $l(t) = a \sqrt{t}$ with 
$a= 6.8 \cdot 10^{-5} \; {\rm m/s^{1/2}}$.

\vspace{0.5cm}
 
\noindent
{\bf Fig. 4.} 
Time evolution of the near membrane layer for two membranes of different
permeabilities: $\circ$ -- membrane 1, $\triangle$ -- membrane 2.
The solid line represents $l(t) = a \sqrt{t}$ with 
$a= 6.8 \cdot 10^{-5} \; {\rm m/s^{1/2}}$.

\end{document}